\documentstyle{amsart}

\newcommand{\C}{{\Bbb C}}
\newcommand{\ac}{{\alpha_{\Bbb C}}}
\newcommand{\R}{{\Bbb R}}
\newcommand{\Z}{{\Bbb Z}}
\newcommand{\M}{{\cal M}}
\newcommand{\Md}{{\cal M}_{\bf d}}
\newcommand{\ra}{{\longrightarrow}}
\newcommand{\F}{{\cal F}}
\newcommand{\Q}{{\Bbb Q}}

\newcommand{\RE}{\operatorname{Re}}
\newcommand{\IM}{\operatorname{Im}}

\newlength{\halfbls}

\newtheorem{Theorem}{Theorem}

\newtheorem{ClassTheorem}{Classification Theorem}

\newtheorem{MainTheorem}{Main Theorem}

\newtheorem{Conjecture}{Conjecture}

\theoremstyle{remark}

\begin{document}

\title
{Lyapunov exponents and Hodge theory}

\author{Maxim Kontsevich}
\address{Institut des Hautes \'Etudes Scientifiques,
Le Bois-Marie, 35 Route de Chartres,
\linebreak
F-91440 Bures-sur-Yvette, France}
\email{maxim@@ihes.fr}

\author{Anton Zorich}
\address{Mathematical College, Independent University of Moscow, Russia
\hspace*{50pt} \linebreak
        %
\hspace*{26pt} and \hspace*{315pt} \linebreak
\hspace*{7.5pt}
Institut Math\'ematique de Rennes,
Universit\'e Rennes-1,
Campus de Beaulieu,
\hspace*{20pt}  \linebreak
35042 Rennes, cedex, France
}
\email{zorich@@univ-rennes1.fr}


\maketitle


\begin{abstract}
Claude  Itzykson  was fascinated  (among  other  things)  by  the
mathematics of  {\it  integrable\rm}  billiards  (see [AI]). This
paper  is devoted  to  new results about  the  {\it chaotic  \rm}
regime. It is an extended version of the talk of one of us (M.K.)
on the  conference  ``The Mathematical  Beauty  of Physics'', Saclay,
June 1996, dedicated to the memory of C.~Itzykson.
\end{abstract}

\vspace*{1cm}

We started  from computer experiments with simple one-dimensional
ergodic dynamical  systems,  and  quite  unexpectedly  ended with
topological string theory. The result  is  a  formula  connecting
fractal dimensions  in  one dimensional ``conformal field theory''
and explicit integrals over certain moduli  spaces.  Also  a  new
analogy  arose  between  ergodic  theory  and  complex  algebraic
geometry.

We will finish the  preface with a brief summary of what  is left
behind the scene. Our moduli spaces are close  relatives of those
arising  in   Seiberg-Witten   approach   to  the  supersymmetric
Yang-Mills theory. The integrals in the main formula  can also be
considered as correlators in a  topological  string  theory  with
$c=1$. Probably, there  is  way to calculate them  in  terms of a
matrix model and an integrable  hierarchy.  In  the derivation we
use some identity in K\"ahler geometry which looks like a  use of
$N=2$ supersymmetry.

\section{Interval exchange transformations}

Let us consider a classical mechanical system with  the action of
a quasi-periodic external force. Mathematically such a system can
be described as a  symplectic  manifold $(X,\omega)$ and a closed
$1$-form  $\alpha$  on  it.  The  Hamiltonian  of  the  system  is  a
multivalued  function  $H$  such  that $dH=\alpha$. Branches  of  $H$
differ from each other by additive constants. One can
 write the equations of motions
   $dF(x(t))/dt =\{F,H\}(x(t)),\, \,\,F\in C^{\infty}(X),$
as  usual.  In   contrast  with  the  case  of  globally  defined
Hamiltonians, the  system  does  not  have  first integrals. More
precisely, one still can make a reduction to codimension $1$ near
local minima  or maxima of  $H$. Nevertheless, the dynamics on an
open part of $X$ is expected to be ergodic.

Many  physical  systems  produce   after   averaging  multivalued
Hamiltonians.  Examples  include  celestial  mechanics,  magnetic
surfaces,  motions  of  charged  particles on Fermi  surfaces  in
crystals, etc. (see the survey of S.P.Novikov [Nov]).

We consider the simplest case of $2$-dimensional phase space.
 Thus we have a closed oriented surface $\Sigma$ with an area
element $\omega\in\Omega^2(\Sigma)$
and an area-preserving vector field $\xi$:
 $$i_{\xi} \omega=\alpha\in \Omega^1(\Sigma),\,\,\,\,d\alpha=0
 ,\,\,\,\,Lie_{\xi}(\omega)=0$$

The main feature of the $2$-dimensional case is that the system
depends essentially on a finite number of parameters.
Generically the surface splits into a finite number of components
filled with periodic trajectories, and finite number of
{\it minimal} components, where every trajectory is dense.
We can associate with  every minimal component a so-called
 {\it interval exchange transformation  \rm} $T$ (see [CFS]).
 First of all, we choose an interval $I$ on $\Sigma$ transversal to
the vector field $\xi$. The transformation $T$  is defined as
the first return map (the Poincar\'e map) from $I$ to itself.
 The form $\alpha$ defines a measure $dx$ and an orientation on
$I$. The map $T$ preserves both $dx$ and the orientation.
 Also, it is easy to see that
 generically $T$ has a finite number of discontinuity points
$a_1,\dots a_{k-1}$ where $k$ is the number of intervals of
continuity of $T$. Thus
 we can identify $I$ with an interval in $\R$ and write $T$
as follows:
 $$I=[0,a]\subset \R,\,\,\,\,0=a_0<a_1<a_2\dots<a_{k-1}<a=
a_k$$
$$T(x)=x+b_i\,\,\,\,for\,\,\,\,\,a_i<x<a_{i+1}$$
where $b_i\in \R,\,\,\,i=0,\dots, k-1$, are some constants.
Moreover, intervals $(a_0,a_1)$, $(a_1,a_2),\dots,(a_{k-1}$, $a_{k})$
 after the application of this map will be situated on $I$ in
an order described by a permutation $\sigma\in S_{k}$
and without overlapping.
Thus
numbers $b_i$ can be reconstructed uniquely from
 the numbers $a_i$ and the permutation $\sigma$.

 We did not use the area element on the
 surface $\Sigma$ in this construction. Everything is
defined in terms
  of a closed $1$-form $\alpha$ and an orientation on $\Sigma$.
Thus it is enough to have an oriented foliation with a
transversal measure (and with finitely many singularities)
on an oriented surface. It is easy to go back from interval exchange
transformations to oriented surfaces with measured foliations.
 The systems which we will get by the inverse construction correspond
 to multivalued Hamiltonians $H$ without local minima and maxima.

Also we can consider
 possibly non-orientable foliations with transversal measures
 on possibly non-orientable surfaces. This leads to the
 consideration of mechanical systems with various additional
 symmetries. The first return map is defined on an interval in an
appropriate ramified double covering of the surface.

The permutation $\sigma$ is called {\it irreducible \rm} if,
 for any $j,\,\,\,1\le j\le k-1$, one has
$$\sigma(\{1,2,\dots,j\})\ne  \{1,2,\dots,j\}$$
It is called {\it nondegenerate} if in addition it obeys some extra
conditions (see 3.1-3.3 in [M] or equivalent conditions 5.1-5.5 in [V1]).
Morally these conditions mean that the permutation does not determine
any fake zeros of the form $\alpha$ --- zeros of order $0$. In particular
$\sigma(j+1)\ne \sigma(j)+1\quad j=1,\dots,k-1$.
First return maps for ergodic flows give irreducible permutations;
appropriate choice of transversal gives nondegenerate permutations.

\begin{Theorem}{\bf (H.~Masur [M]; W.~Veech [V1]).}
Let us consider the interval exchange map $T$  for an irreducible
permutation $\sigma$  and generic values of continuous parameters
$a_i$  (generic  with  respect  to  the  Lebesgue measure on  the
parameter space $\R_+^k=\{(a_1,\dots,a_k)\}$). Then  the  map $T$
is ergodic with respect to the Lebesgue measure $dx$.
\end{Theorem}
The entropy of the map $T$ is $0$.

An analogous result is true for non-orientable measured
foliations on {\it orientable \rm} surfaces. The case of
non-orientable surfaces is always degenerate. In such case
 foliations almost always have  non-trivial families
of closed leaves (see [N]). Presumably, the interesting (ergodic)
 part is always reduced to measured foliations on orientable surfaces.
In order to simplify the exposition we will mainly consider here
 the case when both the surface and the foliation are orientable.

\section{Error terms: first results and computer experiments}

Let $x$ be a generic point on $I$ and let $(y_1,y_2)$ be a
generic subinterval of $I$. Since the map $T$ is ergodic (for generic
 values of lengths of subintervals) we have the following equality
$$\#\{i:0\le i \le N-1,\,\,\,T^{(i)}(x)\in (y_1,y_2)\}\,=\,(y_2-y_1)N+o(N)$$
as $ N \ra +\infty$.
It was first observed in computer experiments (see [Z1]) that this error term (denoted above by $o(N)$)
typically has the growth of a power of $N$,
$$error\,\,\,\,\,term\, \sim O(N^{\lambda})\,\,\,.$$
Here $\lambda<1$ is an universal exponent depending only on the
permutation $\sigma$ (see [Z2] for the proof of the related statement).

In the case of $2$ or $3$ intervals, one has $\lambda=0$. In these cases
 the genus of the surface is $1$ and the transformation itself
 is equivalent to the generic irrational rotation of a circle.

In the case of $4$ intervals for all nondegenerate permutations one has
 $$\lambda=0.33333+ \sim(10^{-6})$$

In the case of $5$ intervals for all nondegenerate permutations one has
$$\lambda=0.50000+\sim(10^{-6})$$

These two cases correspond to surfaces of genus $2$.

If we have $6$ intervals (surfaces of genus $3$),  then the number
$\lambda$ depends on the permutation:
$$\lambda=0.6156\dots \,\,\,\text{or}\,\,\,\,0.7173\dots\,\,\,\,.$$

These two numbers are probably  irrational.

Also computer experiments show (see [Z1]) that a
generic closed $1$-form on a surface $\Sigma$ defines a filtration on
$H_1(\Sigma,\R)$ (``fractal Hodge structure'') by subspaces
$$H_1(\Sigma,\R) \supset F^{\lambda_g}\supset \dots\supset F^{\lambda_2}\supset
F^{\lambda_1}\supset 0,\,\,\,\,g=\,\text{genus of }\Sigma,\,\,\,\,\,
\dim(F^{\lambda_j})=j$$
where    $1=\lambda_1>\lambda_2>\dots>\lambda_g>0$    are    some
universal constants depending only on the permutation. The number
$\lambda$ which  gives the error term  in the ergodic  theorem is
the  second  exponent  $\lambda_2$.  The  highest   term  of  the
filtration   $F^{\lambda_g}$   is  a   Lagrangian   subspace   of
$H_1(\Sigma,\R)$.

One can see numbers $\lambda_i$ geometrically. Let us consider a generic
trajectory of the area-preserving vector field $\xi$ on $\Sigma$. We consider
a sequence of pieces of this trajectory $x(t)$ of lengths $l_j\ra+\infty,\,\,\,
j=1,2,\dots $ such that $x(l_j)$ is close to the starting point $x(0)$. We
connect two ends of these pieces by short intervals
and get a sequence of closed oriented curves $C_j$ on $\Sigma$. Homology
classes of curves $C_j$  are
elements ${\bf v}_j=[C_j]$ in the  group $H_1(\Sigma,\Z)$.

Vectors ${\bf v}_j$ at the first approximation are close to a one-dimensional
space,
  $${\bf v}_j={\bf u}\,l_j
+o(l_j)$$
where ${\bf u}$ is a non-zero element of $H_1(\Sigma,\R)$. Homology class ${\bf u}$ is
  Poincar\'e dual to
the cohomology class $[\alpha]$ of the  $1$-form $\alpha$. The lowest
 non-trivial term in the  filtration is
$$F^{\lambda_1}=\R {\bf u}\,\,\,.$$
 After the projection to  the quotient space
   $H_1(\Sigma,\R)\ra H_1(\Sigma,\R)/\R {\bf u}$
 we get again a sequence of vectors. It turns out that for
 large $j$ these vectors are again close to a $1$-dimensional
 subspace $\cal L$. Also these vectors   mostly will have size
 $(l_j)^{\lambda_2+o(1)}$. We define $2$-dimensional space $F^{\lambda_2}\subset
 H_1(\Sigma,\R)$ as the inverse image of the $1$-dimensional space
 $\cal L$. We can repeat the  procedure $g$ times. On the last
 step we get a chaotic sequence of vectors
 of bounded length in the $g$-dimensional quotient space
 $H_1(\Sigma,\R)/F^{\lambda_g}$ (see also [Z3]).

There  is,  presumably,  an  equivalent way to  describe  numbers
$\lambda_i$. Namely, let $\phi$ be a smooth function on $\Sigma$.
Assume for simplicity that the multi-valued  Hamiltonian has only
non-degenerate (Morse) singularities.   Then,  for  generic   trajectory
$x(t)$, we expect that the  number  $\int_0^T  \phi(x(t)) dt$ for
large $T$ with high probability has size $T^{\lambda_i+o(T)}$ for
some $i\in\{1,\dots,g\}$.  Exponent  $\lambda_1$  appears for all
functions   with   non-zero    average   value,   $$\int_{\Sigma}
\phi\omega\ne 0\,\,\,\,.$$ The next exponent, $\lambda_2$, should
work   for   functions   in   a  codimension  $1$   subspace   in
$C^{\infty}(\Sigma)$ etc.

We discovered  in computer experiments (more
than $100$ cases) that the sum of numbers $\lambda_j$ is  rational,
 $$\lambda_1+\dots+\lambda_g\in \Q\,\,\,.$$

For example, if genus of $\Sigma$ is $3$ and we have $4$ simple saddle
points for the foliation, then
 $$\lambda_1+\lambda_2+\lambda_3=1+0.5517\dots+0.3411\dots=53/28\,\,\,.$$

Also, our observation explains why the case of genus $2$ is exceptional.
If we have two numbers first of which is equal to $1$ and the sum is
rational, then the second number is rational too.

\section{Moduli spaces}

We want to study the renormalization procedure for
interval exchange maps.
 For example, we can define a map from the space of parameters
$$\{(a_1,\dots,a_{k};\sigma)\}=\R_+^{k}\times S_k$$
to itself  considering the first return map of the half $[0,a/2]$ of
the original interval $I=[0,a]$.
 There are also other ways, but the most elegant is the one which we  describe
  at the end of this section. In order to do it we introduce,
  following H.~Masur and W.~Veech
   certain moduli space.

 The space $\Omega^1_{\text{closed}}(\Sigma)/\text{Diff}(\Sigma)$ of equivalence classes of
  closed $1$-forms
on a surface $\Sigma$  is non-Hausdorf.  In order to cure it we  consider
a ``doubling'' of this space consisting of the space of
closed complex-valued $1$-forms $\alpha_{\C}$  satisfying the condition
$$\RE\alpha\wedge
\IM\alpha_{|_x}> 0$$
 for almost all points $x\in \Sigma$. The notion of positivity here is well-defined
because the surface $\Sigma$ is oriented.

  The real-valued $1$-form $\alpha$ whose leaves we considered before
is the real part, $\RE\alpha_{\C}$, of the complex-valued form $\alpha_{\C}$.
  First of all, we should be sure that we did not restrict ourselves.
  It follows from results of E.~Calabi (see [C]), or from results of A.~Katok (see [K]) that,
   for any closed real $1$-form $\alpha$ giving
 a minimal (everywhere dense) foliation,
  there exists at least one closed $1$-form $\alpha'\ne 0$ such that
 $\alpha\wedge\alpha'\ge 0$ everywhere except points where $\alpha$ vanishes.
 Thus we have a complex valued closed $1$-form $\alpha_{\C}=\alpha+i\alpha'$.

 Any such complex-valued $1$-form defines a complex structure
 on $\Sigma$. Locally outside of zeroes of $\ac$ there
is a complex-valued coordinate $z:\Sigma\ra \C$ such that $dz=\ac$.
 Holomorphic functions are defined as continuous functions
 holomorphic in coordinate $z$. Also, there is a canonical
flat metric
 $(\RE\ac)^2+(\IM\ac)^2$ on $\Sigma$ with singularities at zeroes
 of $\ac$.

Let us fix a sequence of non-negative integers
${\bf d}=(d_1,\dots,d_n)$ such that $\sum_i d_i= 2g-2$ where
$g\ge 2$ is the genus of the surface. We denote by
 $\M_{\bf d}$ the moduli space of triples $(C;p_1,\dots,p_n;\ac)$
where $C$ is a smooth complex curve of genus $g$, $p_i$ are
 pairwise distinct points of $C$, and $\ac$ is a holomorphic
 $1$-form on $C$ which vanishes up to order $d_i$ at $p_i$
and is non-zero at all other points of $C$. From this definition
 it is clear that $\M_{\bf d}$ is a Hausdorf complex analytic
(and algebraic) space (see [V3]).

 First of all, $\Md$ is a complex orbifold of dimension
 $2g-1+n$. Let us consider the period map from a neighborhood
of a point $(C;p_1,\dots,p_n;\ac)$ of $\Md$ into the cohomology group
$H^1(C,\{p_1,\dots,p_n\};\C)$.
 Closed form $\ac$ defines an element of the relative
cohomology group  $H^1(C,\{p_1,\dots,p_n\};\C)$ by integration along
paths connecting points $p_i$. In a
neighborhood of any point $(C;p_1,\dots,p_n;\ac)$ of $\Md$, we can
identify cohomology groups $H^1(C',\{p_1',\dots,p_n'\};\C)$ with
 $H^1(C,\{p_1,\dots,p_n\};\C)$
 using the
 Gauss-Manin connection.

 Thus we get a map (the period map) from this neighborhood into a vector space.
 An easy calculation shows that the deformation theory is
not obstructed
 and we get locally a one-to-one correspondence between $\Md$
 and an open domain in the vector space  $H^1(C,\{p_1,\dots
,p_n\};\C)$.

 We claim that $\Md$ has structures $1),2),3),4)$ listed below.

\begin{enumerate}
\item
\it{ a  holomorphic affine structure on $\Md$ modelled on
the vector
space}\linebreak  $H^1(C,\{p_1,\dots,p_n\};\C)$, \rm
\item
\it{ a smooth measure $\mu$ on $\Md$, } \rm
\item
\it{ a locally quadratic non-holomorphic function $A:\Md\ra \R_+$},\rm
\item
\it{a non-holomorphic  action of the group $GL_+(2,\R)$ on $\Md$}.\rm
\end{enumerate}

The first structure we already defined using the period map.

The tangent space to
 $\Md$ at each point contains a lattice,
\begin{multline*}
H^1(C,\{p_1,\dots,p_n\};\C)=H^1(C,\{p_1,\dots,p_n\};\R\oplus
 i\R)\supset \\
\supset H^1(C,\{p_1,\dots,p_n\};\Z)\oplus i \cdot H^1(C,\{p_1,\dots,p_n\};\Z)\,.
\end{multline*}
 The Lebesgue measure (= the Haar measure) on the tangent space to $\Md$ can
be uniquely normalized
  by the condition that the volume of the quotient torus is
 equal to $1$. Thus we defined the density of a measure $\mu$ at
each point of $\Md$.

We define the function $A:\Md\ra\R_+$ by the formula $A(C,\ac)=
{i \over 2}\int_C \ac\wedge\overline{\ac}$. In other terms, it
is the area of $C$ for the flat metric associated with $\ac$.

The group  $GL_+(2,\R)$ of $2\!\times\! 2$-matrices with positive determinant
 acts by linear transformations with
constant coefficients on the pair of real-valued $1$-forms
$\left(\RE(\ac),\IM(\ac)\right)$. In the local affine coordinates,
this action  is the action of
$GL_+(2,\R)$ on the vector space
$$
H^1(C,\{p_1,\dots,p_n\};\C)\simeq \C\otimes H^1(C,\{p_1,\dots,p_n\};\R)
\simeq\R^2 \otimes H^1(C,\{p_1,\dots,p_n\};\R)
$$
 through the first factor in the tensor product.
 From this description it is clear that the subgroup
 $SL(2,\R)$ preserves the measure $\mu$ and the function $A$.

On the hypersurface $\Md^{(1)}=A^{-1}(1)$ (the level set of the
function $A$) we define the induced measure by the formula
$$\mu^{(1)}={\mu\over dA}$$

 The group $SL(2,\R)$ acts on $\Md^{(1)}$ preserving
 $\mu^{(1)}$.

\begin{Theorem}[H.~Masur; W.~Veech]
The total volume of $\Md^{(1)}$ with respect to the measure
  $\mu^{(1)}$ is finite.
\end{Theorem}

Let us denote by $\M$ any connected component of $\Md$ and by $\M^{(1)}$
its intersection with $\Md^{(1)}$.

\begin{Theorem}[H.~Masur; W.~Veech] \
The action of the one-parameter group \linebreak
$\{diag(e^t,e^{-t})\}\subset$ $SL(2,\R) $ on $(\M^{(1)},\mu^{(1)})$
 is ergodic.
\end{Theorem}

The action in this theorem is in fact the renormalization
group flow for interval exchange maps. Another name for this
 flow is the ``Teichm\"uller geodesic flow'', because it gives
 the Euler-Lagrange equations for geodesics for the Teichm\"uller
metric on the moduli space of complex curves. Notice that this
 metric is not a Riemannian metric, but only a Finsler metric.

The intuitive explanation of the ergodicity is that the group
 $\{diag(e^t,e^{-t})\}$ expands leaves of the foliation by affine
 subspaces parallel to
 $H^1(C,\{p_1,\dots,p_n\},\R)$ and contracts leaves of the  foliation by subspaces
 parallel to   $H^1(C,\{p_1,\dots,p_n\},i\R)$.

\section{Topology of the moduli space}

In the last theorem from the previous section
 we consider {\it connected components \rm}
of the moduli space
 $\Md$. From the first glance it seems to be not necessary because
 normally moduli spaces are connected. It is not true in our case.
 W.~Veech and P.~Arnoux discovered by direct calculations in terms of permutations
  that there are several connected components. The set of irreducible
   permutations is decomposed into certain equivalence classes called
extended   Rauzy classes. These classes correspond to connected components of spaces
   $\Md$. For a long
 time the geometric origin of non-connectedness  was not clear.

  Recently we  have obtained the complete classification
  of connected components.
  First of all, there are two series of
  connected components of $\Md$ consisting of hyperelliptic
  curves such that the set of singular points is invariant under
  the hyperelliptic involution. The first series corresponds to curves with
  one singular
   point, ${\bf d}=(2g-2)$ for $g\ge 2$. The second series corresponds to curves
   with  two singular points, ${\bf d}=(g-1,g-1)$.

  If all orders of zeros are even numbers, we have a
  spin structure on $C$ given by a half of the canonical divisor
$$S=\sum_i { d_i \over 2} [p_i]\in Pic(C)\,\,\,.$$

It  is  well-known  that  spin  structures   have  a  topological
characteristic (parity)  which  does  not change under continuous
deformations (see [At]). The {\it parity of a  spin structure} is
the parity of the dimension  of  the space of global sections  of
the corresponding line bundle.

\begin{ClassTheorem}
There   are   hyperelliptic   and   non-hyperelliptic   connected
components  of  the  moduli  spaces of holomorphic  1-forms.  For
non-hyperelliptic components  there  are  two  cases:  the vector
${\bf d}$ is divisible by $2$,  or not. If ${\bf d}$ is divisible
by $2$  then there are two  components corresponding to  even and
odd spin  structures. There are  exceptional cases when we get an
empty set: 1) for $g=2$: all  non-hyperelliptic  strata;  2)  for
$g=3$: non-hyperelliptic strata  with  ${\bf d}$ divisible by $2$
and even spin structure.
\end{ClassTheorem}

We have analogous results for the moduli space of quadratic
differentials.
 At the moment we do not know anything about the topology of
 connected components except for the hyperelliptic locus.

\begin{Conjecture} Each connected component $\M$ of $\Md$ has
homotopy type  $K(\pi,1)$, where $\pi$ is a group commensurable with
some mapping class group.
\end{Conjecture}

\section{Lyapunov exponents}

We recall here the famous  multiplicative  ergodic  theorem  (see
also the  other famous theorem  on the related matter --- theorem
of H.~Furstenberg [F] for a product of random matrices).

\begin{Theorem}{\bf (V.~Oseledets [O]).}
Let $T_t:(X,\mu)\ra (X,\mu),\,\,\,t\in \R_+$, be an ergodic flow
on a space $X$ with finite measure $\mu$; let $V$ be an
$\R_+$-equivariant measurable finite-dimensional real vector bundle.
 We also assume that a (non-equivariant) norm $\|\,\,\,\|$ on $V$ is
 chosen such that, for all $t\in \R_+$,
$$\int_X \log\left(1+\|T_t:V_x\ra V_{T_t(x)}\|\right)\mu<+\infty\ .$$
 Then there are real constants $\lambda_1> \lambda_2>\dots> \lambda_k$ and an
 equivariant filtration of the vector bundle $V$
$$V=V_{\lambda_1}\supset  \dots \supset V_{\lambda_k}\supset 0$$
such that, for almost all $x\in X$ and all $v\in V_x\setminus\{0\}$, one has
 $$\|T_t(v)\|=\exp(\lambda_j t+o(t)),\,\,\,\,\,\,\,\,t\ra +\infty$$
where $j$ is the maximal value for which $v\in (V_{\lambda_j})_x$.
  The filtration $V_{\lambda_j}$ and numbers $\lambda_j$ do not change if we
  replace norm $\|\,\,\,\|$ by another norm $\|\,\,\,\|'$ such that
$$
\int_X \log\left(\max_{v\in V_x\setminus\{0\}}
\left(\max\left( \frac{\|v\|}{\|v\|'},
\frac{\|v\|'}{\|v\|}\right)\right)\right)\,\mu
\ <+\infty \ .$$
\end{Theorem}

Analogous statement is true for systems in discrete time $\Z_+$.

Numbers  $\lambda_j$   are   called  Lyapunov  exponents  of  the
equivariant vector  bundle  $V$.  Usually  people  formulate this
theorem using  language  of matrix-valued $1$-cocycles instead of
equivariant vector bundles. This is equivalent to the formulation
above because any  vector  bundle on  a  measurable space can  be
trivialized on the complement to a subset of measure zero.

If our system is reversible, we can change the positive direction
of  the  time. Lyapunov exponents will be  replaced  by  negative
Lyapunov  exponents.  A new  filtration  will  appear.  This  new
filtration is opposite to the previous  one,  and  they  together
define an equivariant  splitting of $V$  into the direct  sum  of
subbundles.

Lyapunov exponents are, in  general,  very hard to evaluate other
than  numerically.  We  are  aware  only  about  two  examples of
explicit formulas. One example is the geodesic flow  on a locally
symmetric domain and $V$ being  a  homogeneous  vector bundle. In
this case one can explicitly construct the splitting  of $V$. The
second  example  is  the  multiplication  of  random  independent
matrices  whose  entries  are  independent  equally   distributed
Gaussian  random  variables.  In  this  case  one can
calculate Lyapunov  exponents using rotational invariance and the
Markov property.

Our calculation seems to be the  first  calculation  of  Lyapunov
exponents in a  non-homogeneous situation. As the reader will see
later, our proof uses a replacement of a  deterministic system by
a Markov process.

Let us define a vector bundle  $H^1$ over $\M$ by saying that its
fiber at a point $(C;p_1,\dots,p_n;\ac)$ is  the cohomology group
$H^1(C,\R)$. We apply  the  multiplicative ergodic theorem to the
action of $\{diag(e^t,e^{-t})\}$ on $\M^{(1)}$ and  to the bundle
$H^1$. The action of  the group on this bundle is defined  by the
lift using  the natural flat connection (Gauss-Manin connection).
We will  not specify  for a moment the norm  on $H^1$ because all
natural  choices  are equivalent in the sense  specified  in  our
formulation of the multiplicative ergodic theorem.

The  structure  group  of  the   bundle   $H^1$   is  reduced  to
$Sp(2g,\R)\subset GL(2g,\R)$. One can see  easily  that  in  this
case Lyapunov exponents form a symmetric subset of $\R$. Also, in
all experiments we have simple spectrum of Lyapunov  exponents,
i.e. the picture is
$$\lambda_1>\lambda_2>\dots>\lambda_g>
\lambda_{g+1}= -\lambda_g>\dots>\lambda_{2g}=-\lambda_1$$

\begin{Theorem}
  1)  The  lowest Lyapunov  exponent  $\lambda_{2g}=-\lambda_1$  is
  equal  to  $-1$  and  has  multiplicity  one.  The  corresponding
  one-dimensional subbundle is $\RE (\alpha) \R\subset H^1$.
  2) The second Lyapunov exponent $\lambda_2$ governs the error term in the
  ergodic theorem for interval exchange maps.
  3)\,The filtration on $H^1$ related with the positive time dy\-namics
  depends locally only on the cohomology class $[\RE\alpha]\in H^1(\Sigma,\!
  \{p_1,\dots,p_n\};\R)$.
\end{Theorem}

The first part is quite easy. At least, the growth of the
norm for the $1$-dimensional bundle $\RE (\alpha) \R\subset H^1$ is
exponential with the rate $1$.

The second part looks more mysterious. We compare two different
 dynamical systems, the original flow on the surface and
 the renormalization group flow on   the moduli space. The time
 in one system is morally an exponent of the time in another system.
 The technical tool here is a mixture of the ordinary (additive) and
 the multiplicative ergodic theorem
 for an action of the group $\operatorname{Aff}(\R^1)$ of affine transformation of
 line.  We are planning to write in a future a detailed proof;
a rather technical proof of a related statement can be found in [Z2].

The third part is not hard, but surprising. In fact, the positive-time
filtration on $H^1$ coincides with the filtration
for {\it real-valued \rm} closed $1$-forms described in section 2.
Thus it is independent on
the choice of the imaginary part.

In computer experiments we observed that the spectrum of Lyapunov exponents
 is simple.
 In the rest of the paper wee will assume for simplicity that the
 non-degeneracy holds always.
  The general reason to believe in it is that there is no additional symmetry
 in the system which can force the Lyapunov spectrum to be degenerate.

\section{Analogy with the Hodge theory}

We see that our moduli space locally is decomposed into the product of
two manifolds
$$H^1(\dots;\R)\times H^1(\dots;i\R)\,\,\,.$$
 More precisely, we have two complementary subbundles in the tangent
 bundle satisfying the Frobenius integrability condition. This is quite
 analogous the geometry of a complex manifold. If $N$ is an almost complex
 manifold, then we have two  complementary subbundles $T^{1,0}$ and $T^{0,1}$
 in the {\it complexified \rm} tangent bundle $T_N\otimes\C$. The
 integrability condition of the almost-complex structure is equivalent
 to the formal integrability of distributions $T^{1,0}$ and $T^{0,1}$.

Also, if we have a family of complex manifolds $X_b, \,\,\,b\in B$,
parametrized holomorphically by a complex manifold $B$, then for every
integer $k$ we have a holomorphic vector bundle over $B$ with the fiber
$H^k(X_b;\C)$. This bundle carries a natural flat connection and a
holomorphic filtration by subbundles coming from the standard spectral
sequence.

This picture (variations of Hodge structures, see [G])
is parallel to the situation in the multiplicative ergodic
theorem applied to a smooth dynamical system.
 Let $M$ denote the underlying manifold of the system. The tangent
 bundle $T_M$
is an equivariant bundle. Thus, in the case of ergodicity and convergence
of certain integrals we get a canonical measurable splitting of $T_M$ into
the
 direct sum of subbundles indexed by Lyapunov exponents. It is well known
 in many cases (and is expected in general) that
 these subbundles, and also all terms of both filtrations are integrable,
 i.e.
 they are tangent to leaves of (non-smooth) foliations on $M$.
  Two most important foliations (expanding and contracting foliations)
correspond to terms of filtrations associated with
 all positive or all negative exponents. It is known that if the
 invariant measure
 is smooth then the sum of positive
 exponents is equal to the entropy of the system (Pesin formula).

\section{Formula for the sum of exponents }

The main result of our work is an explicit formula for the sum
 of positive Lyapunov exponents $\lambda_1+\dots+\lambda_g$ for the equivariant
 bundle
$H^1$ over the connected component $\M$ of moduli spaces of curves with
holomorphic $1$-forms.

We want to warn the reader that this equivariant bundle
 is not the whole tangent bundle $T_{\M}$. Lyapunov exponents for $T_{\M}$
are calculated in terms of the numbers $\lambda_j$ in the following way:
\begin{multline*}
2>(1+\lambda_2)>(1+\lambda_3)>\dots> (1+\lambda_g) >
\underbrace{1=\dots= 1}_{n-1} >
\\
>(1-\lambda_g)> \dots > (1-\lambda_2) >
0 = 0 >
-(1-\lambda_2)> \dots > -(1-\lambda_g) >
\\
\phantom{\cfrac{1}{1}}
>  \underbrace{-1=\dots= -1}_{n-1} >
-(1+\lambda_g)> -(1+\lambda_{g-1})> \dots > -(1+\lambda_2) > -2
\end{multline*}
Here $n$ is the number of zeros $p_1, \dots, p_n$ of a corresponding
holomorphic 1-form.
The entropy of the
 Teichm\"uller geodesic flow is equal by the Pesin formula to the complex
 dimension of $\M$.
  In short, what we are computing here is more delicate information
  than the
 entropy of the system.

 Hypersurface $\M^{(1)}$ is isomorphic to the quotient space
$\M/\R^*_+$, where $\R^*_+$ is identified with subgroup
$\{diag(e^t,e^t)\}$ of $GL_+(2,\R)$. We denote by $\M^{(2)}$ the
quotient space
$$\M^{(1)}/SO(2,\R)\simeq \M/\C^{\,*}\,\,\,.$$
This space is a complex algebraic orbifold.

Orbits of the group $GL_+(2,\R)$  define a $4$-dimensional foliation on
$\M$. It induces
 a $3$-dimensional foliation on $\M^{(1)}$ by orbits of $SL(2,\R)$,
 and a $2$-dimensional foliation $\F$ on $\M^{(2)}$. Leaves of  $\F$
 are complex curves in $\M^{(2)}$, but the foliation itself is not
 holomorphic.

$3$-dimensional foliation on $\M^{(1)}$ carries a natural transversal
measure. This measure is the quotient of
$\mu^{(1)}$ by the Haar measure on $SL(2,\R)$. The  transversal measure
on $\M^{(1)}$
induces
a transversal measure on $\M^{(2)}$. We have natural orientations on
$\M^{(2)}$ and on leaves of $\F$ arising from complex structures. Thus
we can construct  differential form $\beta$ such that
$$\beta\in \Omega^{\dim_{\R}\M^{(2)}-2}(\M^{(2)}),\,\,\,d\beta=0,\,\,\,\,
Ker\,\beta=\F\,\,\,.$$

The natural projection $\M\ra \M^{(2)}$ is a holomorphic
$\C^{\,*}$-bundle
 with a Hermitean metric given by the function $A$. Thus we
 have a natural curvature form $\gamma_1\in\Omega^2(\M^{(2)}),\,
 \,d\gamma_1=0$ representing the first Chern class $c_1(\M\ra
 \M^{(2)})$. This form is given locally by the formula
  $$\gamma_1={1\over 2 \pi i} \partial\overline{\partial}\, \log
  \left(A(s)\right)$$
where $s$ is a non-zero holomorphic section of the line bundle $\M\ra \M^{(2)}$.

We also have another holomorphic vector bundle on $\M^{(2)}$. The fiber
of this bundle (denoted by $H^{(1,0)}$) is equal to $H^0(C,\Omega^1_C)$, the term of
the Hodge filtration in $H^1\otimes\C$. This holomorphic bundle carries
 a natural hermitean metric coming from the polarization in Hodge theory.
 The formula for the metric is
$$\|\omega\|^2:={1\over 2 \pi i} \int_{C} \omega\wedge  \overline{\omega},
\,\,\,\,\omega \in \Gamma(C,\Omega^1_{hol})$$
This metric defines again a canonical closed $2$-form $\gamma_2$ representing
 the characteristic class $c_1(H^{(1,0)})$.

\begin{MainTheorem} $$\lambda_1+\dots+\lambda_g={\int_{\M^{(2)}} \beta\wedge
\gamma_2 \over
  \int_{\M^{(2)}} \beta \wedge \gamma_1}$$ \par
\end{MainTheorem}
In this formula  we can not go directly to the cohomology, because
the orbifold
 $\M^{(2)}$ is not compact. In order to overcome this difficulty, we
 constructed
 a compactification $\overline{\M}^{(2)}$ of $\M^{(2)}$ with toroidal
 singularities. All three differential forms $\beta,\gamma_1,\gamma_2$
 in the formula seem to be  smooth on
$\overline{\M}^{(2)}$. Both $\gamma_1$ and $\gamma_2$ represent classes
in $H^2(\overline{\M}^{(2)},\Q)$. It seems that $\beta$ also
represents a rational cohomology class although we don't have a proof yet.
In the case of {\it one \rm}   critical point of $1$-form $\alpha$ it is
true because, by  invariance reasons, the form $\beta$ is proportional to a
power of $\gamma_1$.
 Another possible explanation of rationality is that $[\beta]$ is {\it proportional \rm }
  to a rational class because the part of $H^{dim-2}(
  \overline{\M}^{(2)},\R)$ consisting of classes vanishing
  on boundary divisors,
   can be  one-dimensional. In any
  case, we almost explained the rationality of $\sum_{j=1}^g \lambda_j$ observed
  in experiments.

\section{Proof of the formula }

Any leaf of the foliation $\F$ carries a natural  hyperbolic metric.
The generic leaf is a copy of the upper half-plane $SL(2,\R)/SO(2,\R)$.
   We are studying the behavior of the monodromy of the Gauss-Manin
 connection in $H^1$ along {\it a long geodesic going
 in a random direction on a generic leaf of $\F$\rm}. It was
 an old idea of Dennis Sullivan to replace
 the walk along random geodesic by a random walk on the hyperbolic
 plane
 (the Brownian motion). The trajectory of the random walk goes to infinity
 in a random direction with approximately constant speed.

The meaning of the sum $\lambda_1+\dots+ \lambda_g$ is the following. We move using
the  parallel transport a generic Lagrangian subspace $L$ in the fiber of
$H^1$ and
 calculate the average growth of the volume element $L$ associated with the
 Riemannian metric on $L$ induced from the natural metric (polarization) on
  $H^1$.

As we discuss above, we can replace the geodesic flow by the Brownian motion.
 We will approximate the random walk by a sequence of small
 jumps of a fixed length  in random uniformly distributed
 directions on  the hyperbolic plane.

\proclaim Identity. Fix $x\in \M^{(2)}$ and identify the leaf
$\F_x$ of $\F$ passing through $x$ with the model of the
Lobachevsky plane in unit disc $\{z\in\C\,\vert\,\,|z|<1\}$ in such a
way that $x\mapsto z=0$. Also, we trivialize the vector bundle $H^1$
over $\F_x$ using the Gauss-Manin connection. Then, for any Lagrangian
subspace $L\subset H^1_0$, and for any $\epsilon, \,\,0<\epsilon<1$ the following identity holds:
$${1\over 2\pi} \int\limits_0^{2\pi} d\theta \,\,\log\left({
{volume\,\,\,\, on\,\,\,\, } L, {\,\,\, for\,\,\,\, metric\,\,\,\,
in\,\,\,\, } H^1_{\epsilon e^{i\theta}}\over
{volume\,\,\,\, on \,\,\,\,} L,  { \,\,\,for\,\,\,\, metric \,\,\,\,
in \,\,\,\,} H^1_{0}
}\right)=
\int\limits_{{disc}\,\,\,|z|\le \epsilon}
\log\left(\frac{\epsilon^2}{|z|^2}\right)\gamma_2 \,. $$
\par

The idea of the proof of this identity  follows.   Let us choose a locally constant
basis $l_1,\dots,l_g$
 of $L$ and a basis $v_1(z),\dots,v_g(z)$ in $H^{1,0}_z$ depending
 holomorphically on $z\in  \F_x$. Then we have
 $$\|l_1\wedge\dots\wedge l_g\|_z^2=
{(l_1\wedge\dots\wedge l_g\wedge v_1\wedge\dots\wedge v_g)\otimes
  (l_1\wedge\dots\wedge l_g\wedge \overline{v}_1\wedge\dots\wedge
  \overline{v}_g)
 \over(v_1\wedge\dots\wedge v_g\wedge  \overline{v}_1\wedge\dots\wedge
\overline{v}_g)
\otimes(v_1\wedge\dots\wedge v_g\wedge  \overline{v}_1\wedge\dots\wedge
\overline{v}_g)}
$$
where the numerator and the denominator are considered as elements of the
 one dimensional complex vector space $(\wedge^{2g}(H^1_0\otimes \C))^{
 \otimes 2}$.

If we apply the Laplace-Beltrami operator $\Delta=(1/2\pi i)\times \partial_z
\overline{\partial_z}$ to the logarithms of both sides of the
formula from above, we get that
\begin{multline*}
\Delta (\log(\|l_1\wedge\dots\wedge l_g\|_z^2))=\\
=\Delta({holomorphic
\,\,\,\,function})+
\Delta({antiholomorphic\,\,\,\, function}) + \gamma_2\vert_{\F_x}
\end{multline*}
  It implies that the average growth of the volume on $L$ depends not
  on $L$ but only on the position of the point $x\in \M^{(2)}$.
  Because of ergodicity we can average over the invariant
  probability measure $Z^{-1}\times\mu^{(2)}$, where
 $Z=\int_{\M^{(2)}} \mu^{(2)}$ is the total volume of $\M^{(2)}$.
   The invariant measure $\mu^{(2)}$ is proportional to $\beta\wedge\gamma_1$.
 This explains the denominator in the formula for $\lambda_1+\dots+\lambda_g$.

\section{Generalizations }

In our proof we treat the higher-dimensional moduli space $\M^{(2)}$ as
a ``curve with hyperbolic metric''. In general, in many situations
ergodic foliations with transversal measures and certain
differential-geometric structures along leaves can be considered as
virtual manifolds with the same type of geometric structure. Also,
ergodic actions of groups can be considered as virtual discrete subgroups
(Mackey's philosophy) etc.

Our proof works literally in a different situation. Let $C$ be a complex
curve of genus $g>0$ parametrizing polarized abelian varieties $A_x,\,\,x
\in C$ of complex dimension $G$.

We endow $C$ with the canonical hyperbolic metric and consider the geodesic
flow on it. It gives us an ergodic dynamical system. For the equivariant bundle we
will take the symplectic local system $H^1$ over $C$ with fibers
$H^1(A_x,\R)$.
   Again, the sum of positive Lyapunov exponents is rational:
$$\lambda_1+\dots+\lambda_G={deg(H^{1,0})\over 2g-2}$$

\section{Conjectures on the values of Lyapunov exponents}

Fix a collection of integers  ${\bf  d}  =(d_1,\dots,d_n)$,  such
that each $d_i$, $i=1,\dots,n$, is  either  positive,  or  equals
$-1$. Assume that $\sum_i d_i = 4g-4$. We denote by $\cal{Q}_{\bf
d}$ the moduli  space  of triples $(C;p_1,\dots,p_n;q)$ where $C$
is  a  smooth complex  curve  of genus  $g$,  $p_i$ are  pairwise
distinct  points  of  $C$,  and  $q$  is a meromorphic  quadratic
differential on $C$ with the following properties. It has zero of
order $d_i$ at $p_i$ if $d_i>0$, it has a simple pole at $p_i$ if
$d_i=-1$ and it does not have any other zeros or poles on $C$. We
also require that quadratic differential $q$ is not a square of a
holomorphic differential.

H.Masur and J.Smillie showed in [MS]  that  any singularity data
${\bf d}$  satisfying  conditions  above  can  be realized by  a
meromorphic  quadratic  differential  with  the  following   four
exceptions:
$$
{\bf d}\neq\ \ (\ ),\quad (-1,1),\quad (1,3),\quad (4)
$$
${\cal Q}_{\bf d}$ is a Hausdorf complex analytic (and algebraic)
space (orbifold) (see [V3]).

A meromorphic quadratic differential $q$ on a complex curve $C$ determines a
two-sheet cover (or a ramified two-sheet  cover)  $\pi:  \tilde{C}  \to
C$ such  that $\pi^{\ast} q$  becomes a square of a holomorphic
differential on  $\tilde{C}$. Genus $\tilde{g}$  of the minimal cover and
the  singularity  data  $\tilde{{\bf d}}$  of  the   quadratic
differential  $p^{\ast}q$   is   the   same   for  all  quadratic
differentials from $\cal{Q}_{\bf d}$.  By {\it effective genus} we
will call the difference $ g_{\scriptstyle{e\!f\!f}}=\tilde{g}-g $.

\begin{Conjecture}
For  any  moduli space  $\cal{Q}_{\bf d}$  of  meromorphic
quadratic differentials  on  $\C{}P^1$  the  sum  of the Lyapunov
exponents
$\lambda_1 + \dots + \lambda_{g_{\scriptstyle{e\!f\!f}}}$
equals
$$
\lambda_1 + \dots + \lambda_{g_{\scriptstyle{e\!f\!f}}}
\ = \
1+\sum_{i=1}^n \cfrac{\left[\frac{d_i+1}{2}\right]^2}{d_i+2}
\ = \
\cfrac{1}{4}\,\cdot\,\sum\begin{Sb}j\ such\ that\\
d_j\  is\ odd\end{Sb}
\cfrac{1}{d_j+2}
$$
\end{Conjecture}

By definition of the hyperelliptic components ${\cal M}^{\cal H}_{(2g-2)}$
and ${\cal M}^{\cal H}_{(g-1,g-1)}$ of the moduli spaces
${\cal M}_{(2g-2)}$ and ${\cal M}_{(g-1,g-1)}$ there are canonical
isomorphisms:
$$
{\cal M}^{\cal H}_{(2g-2)} = {\cal Q}_{(\underbrace{{\scriptstyle -1,\dots,-1}}_{2g+1},2g-3)} \qquad
{\cal M}^{\cal H}_{(g-1,g-1)} = {\cal Q}_{(\underbrace{{\scriptstyle -1,\dots,-1}}_{2g+2},2g-2)}
$$
Thus Conjecture above gives hypothetical value for the sum of the
Lyapunov exponents for all hyperelliptic components of the moduli
spaces of holomorphic differentials:
$$
\begin{array}{ccccl}
\lambda_1 + \dots + \lambda_g &
\stackrel{?}{=} & \cfrac{g^2}{2g-1} & \text{for } & {\cal M}^{\cal H}_{(2g-2)}
        \\
\lambda_1 + \dots + \lambda_g &
\stackrel{?}{=} & \cfrac{g+1}{2}    & \text{for } & {\cal M}^{\cal H}_{(g-1,g-1)}
\end{array}
$$

As  we  already  mention  above,  the  second  Lyapunov  exponent
$\lambda_2$ is responsible for the deviation from the average for
interval  exchange  transformations  and  for  related  dynamical
systems. Below we present the approximate  values for $\lambda_i$
for  small  genera. One can see in  particular  that  $\lambda_2$
ranges considerably  already  for these few moduli spaces. Still
we wish to believe in the following conjecture.

\begin{Conjecture}
For the hyperelliptic components ${\cal M}^{\cal H}_{(2g-2)}$ and
${\cal M}^{\cal H}_{(g-1,g-1)}$
$$
\lim_{g\to \infty}\lambda_2 = 1
$$
For all other components and other
moduli spaces of holomorphic differentials
$$
\lim_{g\to \infty}\lambda_2 = \frac{1}{2}
$$
\end{Conjecture}

\clearpage  
\appendix
\section{Approximate values of the Lyapunov exponents for small genera}

In   our    experiments    we   used   averaging   over   several
pseudotrajectories of  the  {\it  fast  Rauzy  induction} --- the
discrete analog of  the  Teichm\"uller geodesic flow. The maximal
length  of   trajectories   available   to   us  ($\sim  10^{10}$
iterations) presumably  allows to compute the numbers $\lambda_i$
with approximately $5$ digits of  precision.  This  precision  is
still insufficient  to  determine  corresponding rational numbers
when their denominator is about $\sim 10^3$ or more.

\bigskip 
\bigskip 

\begin{table}[htb]
\caption{Genus $g=3$}
        %
$$
\begin{array}{|c|c||c|c||c|}
\hline
&&\multicolumn{3}{|c|}{}\\

\multicolumn{1}{|c|}{\text{Types}}&
\multicolumn{1}{|c||}{\text{Hyperelliptic}}&
\multicolumn{3}{|c|}{\text{Lyapunov exponents}}\\

\multicolumn{1}{|c|}{\text{of }}&
\multicolumn{1}{|c||}{\text{or spin}}&
\multicolumn{3}{|c|}{\text{}}\\

\cline{3-5}

\multicolumn{1}{|c|}{\text{zeros}}&
\multicolumn{1}{|c||}{\text{structure}}&
\multicolumn{1}{|c|}{\text{}}&
\multicolumn{1}{|c||}{\text{}}&
\multicolumn{1}{|c|}{\text{}}\\

\multicolumn{1}{|c|}{{\bf d}}&
\multicolumn{1}{|c||}{}&
\multicolumn{1}{|c|}{\lambda_2}&
\multicolumn{1}{|c||}{\lambda_3}&
\multicolumn{1}{|c|}{\overset{g}{\underset{j=1}{\sum}} \lambda_j}\\
[-\halfbls] &&&&\\
\hline &&&& \\ [-\halfbls]

(4)      & hyperelliptic   & 0.6156  & 0.1844  & 9/5      \\
[-\halfbls] &&&&\\
\hline &&&& \\ [-\halfbls]
(4)      & odd & 0.4179  & 0.1821  & 8/5      \\
[-\halfbls] &&&&\\
\hline &&&& \\ [-\halfbls]
(1,3)    & -   & 0.5202  & 0.2298  & 7/4      \\
[-\halfbls] &&&&\\
\hline &&&& \\ [-\halfbls]
(2,2)    & hyperelliptic   & 0.6883  & 0.3117  & 4/2      \\
[-\halfbls] &&&&\\
\hline &&&& \\ [-\halfbls]
(2,2)    & odd & 0.4218  & 0.2449  & 5/3      \\
[-\halfbls] &&&&\\
\hline &&&& \\ [-\halfbls]
(1,1,2)  & -   & 0.5397  & 0.2936  & 11/6     \\
[-\halfbls] &&&&\\
\hline &&&& \\ [-\halfbls]
(1,1,1,1)& -   & 0.5517  & 0.3411  & 53/28    \\
[-\halfbls] &&&&\\ \hline
\end{array}
$$
\end{table}

\clearpage   

\begin{table}[h]
\caption{Genus $g=4$}
        %
$$
\begin{array}{|c|c||c|c|c||c|}

\hline
&&\multicolumn{4}{|c|}{}\\

\multicolumn{1}{|c|}{\text{Types}}&
\multicolumn{1}{|c||}{\text{Hyperelliptic}}&
\multicolumn{4}{|c|}{\text{Lyapunov exponents}}\\

\multicolumn{1}{|c|}{\text{of }}&
\multicolumn{1}{|c||}{\text{or spin}}&
\multicolumn{4}{|c|}{\text{}}\\

\cline{3-6}

\multicolumn{1}{|c|}{\text{zeros}}&
\multicolumn{1}{|c||}{\text{structure}}&
\multicolumn{1}{|c|}{\text{}}&
\multicolumn{1}{|c|}{\text{}}&
\multicolumn{1}{|c||}{\text{}}&
\multicolumn{1}{|c|}{\text{}}\\

\multicolumn{1}{|c|}{{\bf d}}&
\multicolumn{1}{|c||}{}&
\multicolumn{1}{|c|}{\lambda_2}&
\multicolumn{1}{|c|}{\lambda_3}&
\multicolumn{1}{|c||}{\lambda_4}&
\multicolumn{1}{|c|}{\overset{g}{\underset{j=1}{\sum}} \lambda_j}\\
[-\halfbls] &&&&&\\
\hline &&&&& \\ [-\halfbls]
(6)      & hyperelliptic   & 0.7375  & 0.4284  & 0.1198 & 16/7  \\
[-\halfbls] &&&&&\\
\hline &&&&& \\ [-\halfbls]
(6)      &even & 0.5965  & 0.2924  & 0.1107 & 14/7  \\
[-\halfbls] &&&&&\\
\hline &&&&& \\ [-\halfbls]
(6)      & odd & 0.4733  & 0.2755  & 0.1084 & 13/7  \\
[-\halfbls] &&&&&\\
\hline &&&&& \\ [-\halfbls]
(1,5)    & -   & 0.5459  & 0.3246  & 0.1297 & 2     \\
[-\halfbls] &&&&&\\
\hline &&&&& \\ [-\halfbls]
(2,4)    &even & 0.6310  & 0.3496  & 0.1527 & 32/15 \\
[-\halfbls] &&&&&\\
\hline &&&&& \\ [-\halfbls]
(2,4)    & odd & 0.4789  & 0.3134  & 0.1412 & 29/15 \\
[-\halfbls] &&&&&\\
\hline &&&&& \\ [-\halfbls]
(3,3)    & hyperelliptic   & 0.7726  & 0.5182  & 0.2097 & 5/2   \\
[-\halfbls] &&&&&\\
\hline &&&&& \\ [-\halfbls]
(3,3)    & -   & 0.5380  & 0.3124  & 0.1500 & 2     \\
[-\halfbls] &&&&&\\
\hline &&&&& \\ [-\halfbls]
%
(1,2,3)  & -   & 0.5558  & 0.3557  & 0.1718 & 25/12 \\
[-\halfbls] &&&&&\\
\hline &&&&& \\ [-\halfbls]
(1,1,4)  & -   & 0.55419 & 0.35858 & 0.15450& ?     \\
[-\halfbls] &&&&&\\
\hline &&&&& \\ [-\halfbls]
(2,2,2)  &even & 0.6420  & 0.3785  & 0.1928 & 166/75\\
[-\halfbls] &&&&&\\
\hline &&&&& \\ [-\halfbls]
(2,2,2)  & odd & 0.4826  & 0.3423  & 0.1749 & 2     \\
[-\halfbls] &&&&&\\
\hline &&&&& \\ [-\halfbls]
(1,1,1,3)& -   & 0.5600  & 0.3843  & 0.1849 & 66/31 \\
[-\halfbls] &&&&&\\
\hline &&&&& \\ [-\halfbls]
(1,1,2,2)& -   & 0.5604  & 0.3809  & 0.1982 &   ?   \\
[-\halfbls] &&&&&\\
\hline &&&&& \\ [-\halfbls]
(1,1,1,1,2)& - & 0.5632  & 0.4032  & 0.2168 &   ?   \\
[-\halfbls] &&&&&\\
\hline &&&&& \\ [-\halfbls]
(1,1,1,1,1,1)&-& 0.5652  & 0.4198  & 0.2403 &   ?   \\
[-\halfbls] &&&&&\\ \hline
\end{array}
$$
\end{table}

\bigskip

\clearpage   


\end{document}